\documentclass[12pt]{article}
\usepackage{amsmath}
\usepackage{graphicx, psfrag, epsf, grffile}
\usepackage{enumerate}
\usepackage{natbib}
\usepackage{url} % not crucial - just used below for the URL 
\usepackage{color}
\usepackage[titletoc,title]{appendix}
\usepackage{setspace,bm} % For double-spacing, URL, and bold font
\usepackage{amsthm}
\usepackage{hhline}
\usepackage{multirow}
\usepackage{algorithm}

\usepackage{etoolbox} % Bibliography underfull/overfull box fix
\apptocmd{\sloppy}{\hbadness 10000\relax}{}{} % Bibliography underfull/overfull box fix

\usepackage{algpseudocode} % algorithm text formatting

\newcommand{\blind}{0}

\begin{document}
\date{}

\def\spacingset#1{\renewcommand{\baselinestretch}%
{#1}\small\normalsize} \spacingset{1}

%%%%%%%%%%%%%%%%%%%%%%%%%%%%%%%%%%%%%%%%%%%%%%%%%%%%%%%%%%%%%%%%%%%%%%%%%%%%%%

\if0\blind
{
  \title{\bf A Repelling-Attracting Metropolis Algorithm for Multimodality}
  \author{Hyungsuk Tak \hspace{.2cm}\\
  Statistical and Applied Mathematical Sciences Institute\\
  \\
  Xiao-Li Meng\hspace{.2cm}\\
    Department of Statistics, Harvard University\\
   \\
    David A. van Dyk\\%, ,  for his comments, discussion on a pseudo-marginal approach
    Statistics Section, Department of Mathematics, Imperial College London}
  \maketitle
} \fi

\if1\blind
{
  \bigskip
  \bigskip
  \bigskip
  \begin{center}
    {\LARGE\bf Title}
\end{center}
  \medskip
} \fi

%\begin{center}
%rev$\_$v6
%\end{center}

\bigskip
\begin{abstract}
Although the Metropolis algorithm is simple to implement, it often has difficulties exploring multimodal distributions. We propose the repelling-attracting Metropolis (RAM) algorithm  that maintains the simple-to-implement nature  of the Metropolis algorithm, but is more likely to jump between modes. The RAM algorithm is  a Metropolis-Hastings algorithm with a proposal that consists of a downhill move in density that aims to make local modes repelling, followed by an uphill move in density that aims to make local modes attracting. The  downhill move is achieved via a reciprocal Metropolis ratio so that the algorithm prefers downward movement. The uphill move does the opposite using the standard Metropolis ratio which prefers upward movement. This down-up movement in density increases the probability of a proposed move to a different mode. Because the acceptance probability of the proposal involves a ratio of intractable integrals, we introduce an auxiliary variable which creates a term in the acceptance probability that cancels with the intractable ratio.  Using several examples, we demonstrate the potential for the RAM algorithm to explore a multimodal distribution more efficiently than a Metropolis algorithm and with less tuning than is commonly required by tempering-based methods. 
\end{abstract}

\noindent%
\vfill

\newpage
\spacingset{1.45} % DON'T change the spacing!
\section{Introduction and overview}\label{sec:intro}

Multimodal distributions are common in statistical applications. However, the Metropolis algorithm \citep{metropolis1953equation}, one of the most widely used Markov chain Monte Carlo (MCMC) methods,  tends to produce Markov chains that do not readily jump between local modes. A popular MCMC strategy for dealing with multimodality is tempering such as parallel tempering \citep{geyer1991}, simulated tempering \citep{geyer1995}, tempered transitions \citep{neal1996}, and equi-energy sampler \citep{kou2006}. Though powerful, these methods typically require extensive tuning.

Building on Metropolis, we construct an alternative multimodal sampler called the repelling-attracting Metropolis (RAM) algorithm, which is essentially as easy to implement as the original Metropolis algorithm.  RAM encourages a Markov chain to jump between modes more frequently than Metropolis, and with  less tuning requirements than  tempering methods. Since  RAM  is more likely to jump between modes than  Metropolis, the proportions of its iterations that are associated with each mode are more reliable estimates of their relative masses.  

\begin{figure}[t!]
\begin{center}
\includegraphics[scale = 0.28]{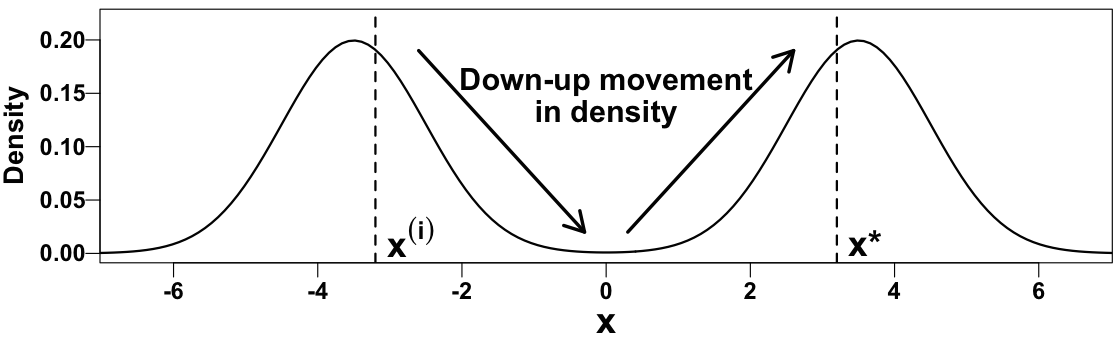}
\end{center}
\caption{A repelling-attracting Metropolis algorithm is a Metropolis-Hastings algorithm that generates a proposal $x^\ast$ given the current state $x^{(i)}$ by making a down-up  movement in density, i.e., repelling-attracting to local modes, via forced downhill and uphill Metropolis transitions.  The  proposal  $x^\ast$ has a higher chance to be near a mode other than the one of the current state, and it is then accepted or rejected in the usual way to preserve the stationary distribution.  }
\label{idea_FM}
\end{figure}

RAM  generates a proposal via forced downhill and forced uphill Metropolis transitions. The term \emph{forced} emphasizes that neither Metropolis transition is allowed to stay at its current state because we repeatedly make proposals  until one is accepted. The forced downhill Metropolis transition uses a reciprocal ratio of the target densities in its acceptance probability. This encourages the intermediate proposal to prefer downward moves since a lower density state has a higher chance of being accepted, hence local modes become \emph{repelling}. The subsequent forced uphill Metropolis transition  generates a final proposal with a standard Metropolis ratio that makes local modes \emph{attracting}. Together the downhill and uphill transitions form a proposal for a Metropolis-Hastings  (MH) sampler \citep{hastings1970monte}, as shown in Fig.~\ref{idea_FM}; a final accept-reject step preserves the stationary distribution. 

As with other MH samplers, the normalizing constant of the target density need not be known, but the scale of the (symmetric) jumping rules used within the downhill and uphill  transitions needs to be tuned. In principle, RAM is designed to improve Metropolis' ability to jump between modes using the same jumping rule as Metropolis where this jumping rule is tuned to optimize the underlying Metropolis sampler for the multimodal target. One could do still better with additional tuning of RAM, but in our experience even with no additional tuning, RAM can perform better than its underlying Metropolis sampler.

Although we can draw a sample using the down-up jumping rule,  the overall acceptance probability contains a ratio of intractable integrals. We can avoid evaluating this ratio by introducing an auxiliary variable \citep{moller2006}. This    preserves the target marginal distribution and requires another forced downhill Metropolis transition for the auxiliary variable. Thus,  RAM  generates a proposal via three forced Metropolis transitions but accepts the proposal with an easy-to-compute acceptance probability.

RAM is related to a number of existing algorithms. The down-up proposal of RAM may be viewed as a simpler version of a mode-jumping proposal \citep{tjelmeland2001mode} whose uphill movement is achieved by a deterministic optimizer. Also, the forced Metropolis transition of RAM is similar to the delayed rejection method \citep{tierney1999some, trias2009delayed} in that both generate proposals repeatedly until one is accepted. RAM's forced transition is a special case of the delayed rejection method in that RAM uses the same jumping rule throughout while  delayed rejection  allows different jumping rules.

In a series of four numerical examples, we compare RAM's performance to Metropolis and  commonly used tempering-based methods such as the equi-energy sampler, parallel tempering, and tempered transitions. We adjust for the required number of evaluations of the target density or the overall CPU time required by each sampler. Our examples range from relatively simple and high dimensional Gaussian mixtures (Examples 1 and 2)  to lower dimensional, but more complex targets that arise as posterior distributions in scientific problems (Examples 3 and 4). We compare RAM with standard Metropolis, implementing both samplers with a common jumping rule that is tuned to improve the mixing of Metropolis for the particular multimodal target distribution.  These  comparisons suggest that replacing Metropolis with RAM when targeting a multimodal distribution can be an efficient strategy, in terms of user's effort. 

In our comparisons with tempering-based samplers, we find that in moderate dimensions RAM performs as well as or better than tempering-based methods, without the subtle tuning that these methods require. Even with a higher dimensional target distribution in Example 3, we show how RAM can be embedded within a Gibbs sampler to obtain results as good as  tempering-based methods, again without the tuning they require. Because RAM is able to jump between modes relatively often, it provides  good estimates of the relative size of the modes. In our examples RAM obtains  more reliable estimates of the mode sizes than Metropolis and is  easier to directly implement than tempering-based methods.

\section{A repelling-attracting Metropolis algorithm}

\subsection{A down-up proposal}
We briefly review  MH. A transition kernel on $\mathbf{R}^d$, denoted by $P(B\mid x)$, is the conditional probability distribution  of a transition from $x\in \mathbf{R}^d$ to a point in a Borel set $B$ in $\mathbf{R}^d$. Hence ${P(\mathbf{R}^d\mid x)=1}$ and $P( \{x\}\mid x)$ need not be zero \citep{chib1995}. A jumping density given the current state $x^{(i)}$ is the conditional  density with respect to Lebesgue measure that generates a proposal $x^\ast$, denoted by  $q(x^\ast\mid x^{(i)})$.   With a target density denoted by $\pi$, either normalized or unnormalized, the transition kernel of  MH   is 
\begin{equation}\label{MH_kernel}
P(dx^\ast\mid x^{(i)})=q(x^\ast\mid x^{(i)})\alpha(x^\ast\mid x^{(i)})dx^\ast +\delta_{x^{(i)}}(dx^\ast)\{1-A(x^{(i)})\},
\end{equation}
where the Dirac measure $\delta_{x^{(i)}}(dx^\ast)$ is one if $x^{(i)}\in dx^\ast$ and zero otherwise and $\alpha(x^\ast\mid x^{(i)})$ is the probability of accepting the proposal and  setting $x^{(i+1)}=x^\ast$, i.e., 

\begin{equation}%\label{accept_mh}
\alpha(x^\ast\mid x^{(i)})=\textrm{min}\left\{1,~ \frac{\pi(x^\ast)q(x^{(i)}\mid x^\ast)}{\pi(x^{(i)})q(x^\ast\mid x^{(i)})}\right\}.\nonumber
\end{equation} 
Here, $1-A(x^{(i)})$ is the probability of staying at $x^{(i)}$, i.e., of setting $x^{(i+1)}=x^{(i)}$, and thus $A(x^{(i)})$ is the probability of moving away from $x^{(i)}$:
\begin{equation}%\label{prob_staying_current}
A(x^{(i)})= \int q(x^\ast\mid x^{(i)})\alpha(x^\ast\mid x^{(i)})dx^\ast.\nonumber
\end{equation}
If the jumping density is symmetric, i.e., $q(a\mid b)=q(b\mid a)$, MH  reduces to  Metropolis  with 
\begin{equation}\label{accept_m}
\alpha(x^\ast\mid x^{(i)})=\textrm{min}\left\{1,~ \frac{\pi(x^\ast)}{\pi(x^{(i)})}\right\}.
\end{equation} 
We assume that $q$ is symmetric  hereafter because RAM is currently feasible only with a symmetric $q$, i.e., RAM can replace any Metropolis but not the more general MH algorithm.

Metropolis  is one of the most commonly used MCMC methods, but it often has difficulties exploring multimodal distributions. Alternative tempering methods usually require more tuning, which can be restrictive to practitioners.  RAM  maintains the simple-to-implement nature of  Metropolis, but is more likely to jump between modes. The key to  RAM  is a down-up jumping density that generates a proposal $x^\ast$ after making a down-up movement in density. Because the corresponding acceptance probability is intractable, we  generate an auxiliary variable $z^\ast$ given $x^\ast$ in such a way that the acceptance probability becomes computable. Thus, RAM is an {MH} algorithm with a unique joint jumping density $q^\textrm{DU}(x^\ast\mid x^{(i)})q^\textrm{D}(z^\ast\mid x^\ast)$ and an easy-to-compute  acceptance probability $\alpha^\textrm{J}(x^\ast, z^\ast\mid x^{(i)}, z^{(i)})$ that preserves the target marginal distribution $\pi(x)$. Next, we describe $q^\textrm{DU}$, $q^\textrm{D}$, and $\alpha^\textrm{J}$.

The  down-up jumping density, $q^{\textrm{DU}}(x^\ast\mid x^{(i)})$, first generates an intermediate downhill proposal $x'$ given the current state $x^{(i)}$ and then an uphill proposal $x^\ast$ given $x'$, i.e.,
\begin{equation}\nonumber%\label{transition_VMH}
q^{\textrm{DU}}(x^\ast\mid x^{(i)}) = \int q^{\textrm{D}}(x'\mid x^{(i)}) q^{\textrm{U}}(x^\ast\mid x')dx',  
\end{equation}
where $q^{\textrm{D}}$ and $q^{\textrm{U}}$ can be any conditional density functions that prefer lower and higher density states than the given  states, respectively. Our choice for $q^{\textrm{D}}$ is a forced downhill Metropolis kernel density defined as
\begin{equation}
q^{\textrm{D}}(x'\mid x^{(i)})=\frac{q(x'\mid x^{(i)}) \alpha^{\textrm{D}}_\epsilon(x'\mid x^{(i)})}{A^\textrm{D}(x^{(i)})},\label{downhill}
\end{equation}
where
\begin{equation}
\alpha^{\textrm{D}}_\epsilon(x'\mid x^{(i)})=\textrm{min}\left\{1,~ \frac{\pi(x^{(i)})+\epsilon}{\pi(x')+\epsilon}\right\}\label{downhill_accept_prob}
\end{equation}
is the  probability of accepting an intermediate proposal $x'$ generated from $q(x'\mid x^{(i)})$ and $A^\textrm{D}(x^{(i)})=\int q(x'\mid x^{(i)}) \alpha^{\textrm{D}}_\epsilon(x'\mid x^{(i)})dx'$ is the normalizing constant\footnote{This normalizing constant $A^\textrm{D}(x^{(i)})$ is finite if $q$ is a proper density, i.e., $\int q(x'\mid x^{(i)}) dx'<\infty$, because $\alpha^{\textrm{D}}_\epsilon(x'\mid x^{(i)})$ is bounded between 0 and 1. Similarly, $A^\textrm{U}(x')$ appearing later is also finite if $q$ is proper.}. We use the term \emph{forced} because this Metropolis transition kernel  repeatedly generates intermediate proposals (like rejection sampling) until one is accepted.  Also we use the term \emph{downhill} because the reciprocal of the ratio of the target densities in \eqref{downhill_accept_prob} makes local modes repelling rather than attracting: If the density of $x'$ is smaller than that of $x^{(i)}$,  $x'$  is accepted with probability one.  The appearance of $\epsilon$ in $\alpha^{\textrm{D}}_\epsilon(x'\mid x^{(i)})$ is discussed below.

Similarly, we set $q^{\textrm{U}}$ to a forced uphill Metropolis transition kernel density defined as 
\begin{equation}
q^{\textrm{U}}(x^\ast\mid x') = \frac{q(x^\ast\mid x') \alpha^{\textrm{U}}_\epsilon( x^\ast\mid x')}{A^\textrm{U}(x')},\nonumber
\end{equation}
where
\begin{equation}
\alpha^\textrm{U}_\epsilon(x^\ast\mid x')=\textrm{min}\left\{1,~ \frac{\pi(x^\ast)+\epsilon}{\pi(x')+\epsilon}\right\}\label{uphill}
\end{equation}
is the  probability of accepting a   proposal $x^\ast$ generated from $q(x^\ast\mid x')$ and \hbox{$A^\textrm{U}(x')=$}\break $\int q(x^\ast\mid x') \alpha^{\textrm{U}}_\epsilon(x^\ast\mid x')dx^\ast$ is the normalizing constant. This  kernel restores the attractiveness of local modes because $\alpha^\textrm{U}_\epsilon(x^\ast\mid x')$ is a typical Metropolis acceptance probability except that $\epsilon$ is added for numerical stability; both $\pi(x')$ and $\pi(x^\ast)$ can be nearly zero  when  both $x'$ and $x^\ast$ are in a  valley between  modes. The value of $\epsilon$ may affect the convergence rate. To minimize its impact on the acceptance probability in~\eqref{uphill}, we choose $\epsilon$ to be small with a default choice of $\epsilon=10^{-308}$, the smallest power of ten that \textrm{R} \citep{r2014} treats as positive. For symmetry, we use $\epsilon$ in the same way in the acceptance probability of the downhill transition  in \eqref{downhill_accept_prob}.  Consequently, our choices for  $q^{\textrm{D}}$ and $q^{\textrm{U}}$  satisfy $\int q^{\textrm{DU}}(x^\ast\mid x^{(i)}) dx^\ast=1$.

Without forced transitions, the final proposal $x^\ast$ could be the same as the current state $x^{(i)}$ after  consecutive  rejections in both the downhill and uphill Metropolis transitions, or  $x^\ast$ could be generated via only one of the downhill and uphill transitions if the other were rejected. This would not be helpful for our purposes because it would not induce a down-up movement. Moreover, a forced transition kernel is mathematically simpler than that of  Metropolis  in that it eliminates the term, $\delta_{x^{(i)}}(dx^\ast)\{1-A(x^{(i)})\}$ in~\eqref{MH_kernel}.

The  MH acceptance probability with the down-up jumping density $q^{\textrm{DU}}$ simplifies to
\begin{equation}
\alpha^{\textrm{DU}}(x^\ast\mid x^{(i)})=\textrm{min}\left\{1,~ \frac{\pi(x^\ast)q^\textrm{DU}(x^{(i)}\mid x^\ast)}{\pi(x^{(i)})q^\textrm{DU}(x^\ast\mid x^{(i)})}\right\}=\min\left\{1,~ \frac{\pi(x^\ast)A^\textrm{D}(x^{(i)})}{\pi(x^{(i)})A^\textrm{D}(x^\ast)}\right\}\label{final_acceptance},
\end{equation}
where the last equality holds because
\begin{align}
q^{\textrm{DU}}(x^\ast\mid x^{(i)})A^\textrm{D}(x^{(i)})&=\int  q(x'\mid x^{(i)}) \alpha^{\textrm{D}}_\epsilon(x'\mid x^{(i)}) \frac{q(x^\ast\mid x')\alpha^{\textrm{U}}_\epsilon(x^\ast\mid x')}{A^\textrm{U}(x')} dx'\nonumber\\
= \int q&(x^{(i)}\mid x') \alpha^{\textrm{U}}_\epsilon(x^{(i)}\mid x') \frac{q(x'\mid x^\ast) \alpha^{\textrm{D}}_\epsilon(x'\mid x^\ast)}{A^\textrm{U}(x')} dx'=q^{\textrm{DU}}(x^{(i)}\mid x^\ast)A^\textrm{D}(x^\ast),\nonumber
\end{align}
and thus
\begin{equation}\label{David_add}
\frac{q^\textrm{DU}(x^{(i)}\mid x^\ast)}{q^\textrm{DU}(x^\ast\mid x^{(i)})}=\frac{A^\textrm{D}(x^{(i)})}{A^\textrm{D}(x^\ast)}.
\end{equation}

\subsection{An auxiliary variable approach}\label{deriv}
Since the ratio of the normalizing constants in~\eqref{David_add} is intractable, we use an auxiliary variable approach \citep{moller2006} to avoid its evaluation  in \eqref{final_acceptance}. We form a joint Markov chain for $x$ and an auxiliary variable $z$ so that the target marginal density for $x$ is still $\pi$,  yet the resulting joint MH algorithm has an easily computable acceptance ratio. Specifically, after generating $x^\ast$ via $q^{\textrm{DU}}$,  we generate $z^\ast$ given $x^\ast$ using the forced downhill Metropolis kernel density $q^{\textrm{D}}$ in \eqref{downhill}, which typically requires one evaluation of the target density on average. We set the joint target density $\pi(x, z)=\pi(x)q(z\mid x)$, which then leads to, as we shall prove shortly, the  acceptance probability of the joint jumping density  $q^{\textrm{DU}}(x^\ast\mid x^{(i)}) q^{\textrm{D}}(z^\ast\mid x^\ast)$ as
\begin{equation}
\alpha^\textrm{J}(x^\ast, z^\ast\mid x^{(i)}, z^{(i)})
=\min\left\{1,~ \frac{\pi(x^\ast)\min\{1, \frac{\pi(x^{(i)})+\epsilon}{\pi(z^{(i)})+\epsilon}\}}{\pi(x^{(i)})\min\{1, \frac{\pi(x^\ast)+\epsilon}{\pi(z^\ast)+\epsilon}\} }\right\}.\label{joint_accept_prob2}
\end{equation}
Consequently, introducing $z$ results in the easy-to-compute acceptance probability in \eqref{joint_accept_prob2}. RAM accepts the joint proposal $(x^\ast, z^\ast)$ as $(x^{(i+1)}, z^{(i+1)})$ with the probability in \eqref{joint_accept_prob2} and sets $(x^{(i+1)}, z^{(i+1)})$ to $(x^{(i)}, z^{(i)})$ otherwise.  Since RAM is an MH algorithm, it automatically satisfies the detailed balance condition.  We notice that in \eqref{joint_accept_prob2}, $\pi(z^{(i)})$ is likely to be smaller than $\pi(x^{(i)})$ because $z^{(i)}$ is generated by the forced downhill  transition. Similarly, $\pi(z^\ast)$ is likely to be smaller than $\pi(x^\ast)$. When $z^{(i)}$ and $z^\ast$ have lower target densities than $x^{(i)}$ and $x^\ast$, respectively (likely, but not required),  the acceptance probability in \eqref{joint_accept_prob2}  reduces  to the acceptance probability of  Metropolis  in \eqref{accept_m}. 

%starting from 
We obtained \eqref{joint_accept_prob2} by considering a joint target distribution $\pi(x, z)=\pi(x)\pi^\textrm{C}(z\mid x)$, with a joint jumping density in the form of 
\begin{equation}\label{factor_joint_prop}
q^\textrm{J}(x^\ast, z^\ast\mid x^{(i)}, z^{(i)})=q_1(x^\ast \mid  x^{(i)}, z^{(i)})q_2(z^\ast \mid  x^\ast, x^{(i)}, z^{(i)})=q_1(x^\ast \mid  x^{(i)})q_2(z^\ast \mid  x^\ast).%\nonumber
\end{equation}
The MH acceptance probability for the joint proposal then is
\begin{align}
\alpha^\textrm{J}(x^\ast, z^\ast\mid x^{(i)}, z^{(i)})=\min\left\{1,~ \frac{\pi(x^\ast)\pi^\textrm{C}(z^\ast\mid x^\ast)q_1(x^{(i)} \mid  x^\ast)q_2(z^{(i)} \mid  x^{(i)})}{\pi(x^{(i)})\pi^\textrm{C}(z^{(i)}\mid x^{(i)})q_1(x^\ast \mid  x^{(i)})q_2(z^\ast \mid  x^\ast)}\right\},\label{joint_accept_prob}
\end{align}
which recalls the pseudo-marginal approach \citep{beaumont2003, andrieu2009} that uses an unbiased estimator of an intractable target density. In this setting, however, it is the jumping density that is intractable. Somewhat surprisingly, there does not seem to be an easy way to modify the pseudo-marginal argument, other than directly following  the more general auxiliary variable approach in \cite{moller2006}.

Specifically, suppose we are able to sample from $q_1$ in \eqref{factor_joint_prop} but are not able to evaluate $q_1$. We can find a function $f$ such that $q_1(x^{(i)} \mid  x^\ast)/q_1(x^\ast \mid  x^{(i)}) = f(x^{(i)})/ f(x^\ast)$ because the ratio of two (compatible) conditional densities equals the corresponding ratio of  marginal densities, where $f$ itself may or may not be computable. If we can find a function $q_2$ in~\eqref{factor_joint_prop} whose normalizing constant is proportional to $f$, then the joint acceptance probability in~\eqref{joint_accept_prob} becomes free of the intractable quantities.

For  RAM, we set $q_1(x^\ast \mid  x^{(i)})=q^\textrm{DU}(x^\ast\mid x^{(i)})$, and thus $f(x^{(i)})=A^\textrm{D}(x^{(i)})$. To eliminate this intractable normalizing constant, we choose $q_2(z^\ast\mid  x^\ast)=q^\textrm{D}(z^\ast\mid x^\ast)$. Since \cite{moller2006} suggest choosing $\pi^\textrm{C}$ similar to $q_2$, we set $\pi^\textrm{C}(z^\ast\mid x^\ast)=q(z^\ast\mid x^\ast)$. With these choices, the acceptance probability in  \eqref{joint_accept_prob} reduces to \eqref{joint_accept_prob2} because
\begin{align}
\alpha^\textrm{J}(x^\ast, z^\ast\mid x^{(i)}, z^{(i)})&=\min\left\{1,~ \frac{\pi(x^\ast)q(z^\ast\mid x^\ast)q^\textrm{DU}(x^{(i)}\mid x^\ast)q^\textrm{D}(z^{(i)}\mid x^{(i)})}{\pi(x^{(i)})q(z^{(i)}\mid x^{(i)})q^\textrm{DU}(x^\ast\mid x^{(i)})q^\textrm{D}( z^\ast\mid x^\ast)}\right\}\nonumber\\
=\min&\left\{1,~ \frac{\pi(x^\ast)q(z^\ast\mid x^\ast)A^\textrm{D}(x^{(i)})q(z^{(i)}\mid x^{(i)}) \alpha^{\textrm{D}}_\epsilon(z^{(i)}\mid x^{(i)})/A^\textrm{D}(x^{(i)}) }{\pi(x^{(i)})q(z^{(i)}\mid x^{(i)})A^\textrm{D}(x^\ast)q(z^\ast\mid x^\ast) \alpha^{\textrm{D}}_\epsilon(z^\ast\mid x^\ast)/A^\textrm{D}(x^\ast)}\right\}\nonumber\\
=\min&\left\{1,~ \frac{\pi(x^\ast)\alpha^{\textrm{D}}_\epsilon(z^{(i)}\mid x^{(i)})  }{\pi(x^{(i)}) \alpha^{\textrm{D}}_\epsilon(z^\ast\mid x^\ast) }\right\}=\min\left\{1,~ \frac{\pi(x^\ast)\min\{1, \frac{\pi(x^{(i)})+\epsilon}{\pi(z^{(i)})+\epsilon}\}}{\pi(x^{(i)})\min\{1, \frac{\pi(x^\ast)+\epsilon}{\pi(z^\ast)+\epsilon}\} }\right\},\nonumber
\end{align}
where the second equality follows from \eqref{downhill} and \eqref{David_add}, and the last equality follows from~\eqref{downhill_accept_prob}.

\subsection{Implementation of the RAM algorithm}\label{implementation_ram}

\begin{table}[b!]
\caption{A repelling-attracting Metropolis algorithm.} \label{al1}
\begin{tabular}{l}
\hline
Set initial values $x^{(0)}$ and $z^{(0)}~(=x^{(0)})$. For $i=0, 1, \ldots$\\
\emph{Step} 1: ($\searrow$) Repeatedly sample $x'\sim q(x'\mid x^{(i)})$ and  $u_1\sim \textrm{Uniform}(0, 1)$\\
~~~~~~~~~~ until $u_1< \textrm{min}\!\left\{1,~ \frac{\pi(x^{(i)})+\epsilon}{\pi(x')+\epsilon}\right\}$.\\
\emph{Step} 2: ($\nearrow$) Repeatedly sample $x^\ast\sim q(x^\ast\mid x')$ and $u_2\sim \textrm{Uniform}(0, 1)$\\
~~~~~~~~~~  until $u_2<\textrm{min}\!\left\{1,~ \frac{\pi(x^\ast)+\epsilon}{\pi(x')+\epsilon}\right\}$.\\
\emph{Step} 3: ($\searrow$) Repeatedly sample $z^\ast\sim q(z^\ast\mid x^\ast)$ and $u_3\sim \textrm{Uniform}(0, 1)$\\
~~~~~~~~~~  until $u_3<\textrm{min}\!\left\{1,~ \frac{\pi(x^\ast)+\epsilon}{\pi(z^\ast)+\epsilon}\right\}$.\\
\emph{Step} 4: Set $(x^{(i+1)}, z^{(i+1)})=(x^\ast, z^\ast)$ if $u_4< \min\left\{1,~ \frac{\pi(x^\ast)\min\{1, (\pi(x^{(i)})+\epsilon)/(\pi(z^{(i)})+\epsilon)\}}{\pi(x^{(i)})\min\{1, (\pi(x^\ast)+\epsilon)/(\pi(z^\ast)+\epsilon)\} }\right\}$, \\
$~~~~~~~~~~$where $u_4\sim  \textrm{Uniform}(0, 1)$, and  set $(x^{(i+1)}, z^{(i+1)})=(x^{(i)}, z^{(i)})$ otherwise.
\end{tabular}
\end{table}

Each RAM iteration  is composed of the four steps in  Table~\ref{al1}. The first three  generate a joint proposal, $(x^\ast, z^\ast)$, via  three consecutive forced transitions; \emph{Step}~1 is  the downward proposal $x'$ given $x^{(i)}$, \emph{Step}~2 is the upward proposal $x^\ast$ given $x'$, and \emph{Step}~3 is the downward proposal $z^\ast$ given $x^\ast$. Finally, \emph{Step}~4 determines if the joint proposal is accepted. In our numerical examples, the downhill proposals in \emph{Steps}~1 and 3 are usually accepted on the first try. However, the number of  proposals needed for the uphill move in \emph{Step}~2 varies. As the dimension increases, for instance, generating a higher density proposal  becomes challenging, and  the uphill transition in \emph{Step}~2 requires  more  proposals.

Some density values used by RAM do not need to be calculated repeatedly. For example, since the density of the previous value $\pi(x^{(i)})$ is  used  in both \emph{Steps}~1 and 4, it is better to evaluate and cache this value before \emph{Step}~1. Also, $\pi(x')$  in \emph{Step}~2 is  evaluated during the final forced downhill step in \emph{Step}~1, and can be cached and  reused in \emph{Step}~2. Similarly, we can cache the values of $\pi(x^\ast)$ and $\pi(z^\ast)$ in \emph{Steps}~2 and 3, respectively. The cached values can also be used to compute the acceptance probability in \emph{Step}~4. In our numerical illustrations, we use an equivalent caching policy for other algorithms. For example, an $\textrm{MH}$ transition can be efficiently coded to evaluate the target density once at each iteration  (only for the density of a proposal) by caching the density of the current state.

RAM can replace a Metropolis kernel within a Gibbs sampler. Suppose we have a Gibbs sampler that iteratively samples $\pi_1(x\mid y)$ and $\pi_2(y\mid x)$, and a Metropolis kernel that is invariant to $\pi_1(x\mid y)$ is used within the Gibbs sampler.  To replace  Metropolis  with RAM, we  keep track of the auxiliary variable $z$  during the run. For example, once we sample $x^{(i)}$ and $z^{(i)}$  at iteration $i$ via a RAM kernel  that is (marginally) invariant to $\pi_1(x\mid y^{(i-1)})$, only $x^{(i)}$ is used to sample $\pi_2(y\mid x^{(i)})$, but $z^{(i)}$ is used  to sample $x^{(i+1)}$ in the next iteration.

For simplicity, we use Gaussian jumping rules,  though any symmetric density can be used. Specifically, we consider a $d$-dimensional Gaussian density with covariance matrix $\Sigma$ as $q$ in Table~1; both RAM and Metropolis share the same tuning parameter $\Sigma$. RAM is designed to improve the ability of Metropolis to jump between modes using a  jumping rule that is tuned to optimize Metropolis  for the multimodal target. In practice, this means a large jumping scale for unknown mode locations or a properly adjusted jumping scale for known mode locations. One could do still better with additional tuning of RAM. For example, if $\Sigma$ is tuned to optimize Metropolis for a multimodal target, we can simply set  the covariance matrix of $q$ for RAM to $\Sigma/2$ because RAM's down-up proposal is generated by two  (down-up) Metropolis transitions.  In our numerical illustrations we show that RAM can improve on Metropolis even without additional tuning. We introduce several useful strategies for tuning $\Sigma$, but their effectiveness may vary in different settings.

\section{Numerical illustrations}\label{examples}

\subsection{Example 1: A mixture of twenty bivariate Gaussian densities}\label{ch3_example1}
%For a comparison with other 

\begin{figure}[b!]
\begin{center}
\includegraphics[scale = 0.3]{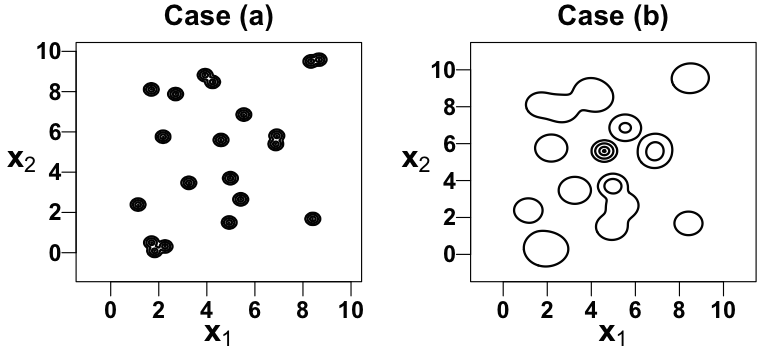}
\caption{The first panel exhibits the contour plot of the target density in Example~1, case~(a) and the second panel shows that of the target density in Example~1, case (b). The plotted contours outline regions with probability 1\%, 10\%, 50\%, and 95\% under $\pi(x)$.}
\label{ee_contour}
\end{center}
\end{figure}

To compare RAM with tempering methods, our first numerical illustration targets a mixture of twenty bivariate Gaussian distributions given in  \cite{kou2006}:
\begin{equation}
\pi(x)\propto \sum_{j=1}^{20}\frac{w_j}{\tau^2_j}\exp\left(-\frac{1}{2\tau^2_j}(x-\mu_j)^\top(x-\mu_j)\right),\nonumber
\end{equation}
where $x=(x_1, x_2)^\top$. The twenty mean vectors,  $\{\mu_1, \ldots, \mu_{20}\}$,  are specified in \cite{kou2006} and plotted in the first panel of Fig.~\ref{ee_contour}. Following \cite{kou2006}, we consider two cases; in case (a), the modes are equally weighted and have equal variances, $w_j=1/20$ and  $\tau^2_j=1/100$, and in case (b) weights and variances are unequal, $w_j=1/\lVert \mu_j- (5, 5)^\top\rVert$ and {$\tau^2_j=\lVert \mu_j-(5, 5)^\top\rVert/20$}. In case (b), modes near (5, 5) have higher weight and smaller variances.  Contour plots of the target distributions in cases (a) and (b) appear in Fig.~\ref{ee_contour}.

\cite{kou2006} used this target distribution to compare the equi-energy sampler (EE) and parallel tempering (PT). We  follow their simulation configurations by running  RAM  for 75,000 iterations for both cases, initializing the chain at random values of $x^{(0)}$ and $z^{(0)}$ in the unit square. Although \cite{kou2006} do not specify the burn-in size, we discard the first 25,000 iterations because they consistently use one third of the iterations as burn-in in the other examples. We set $q$ to be Gaussian with  covariance matrix $\sigma^2I_2$, where $I_2$ is the identity matrix. To tune $\sigma$, we initilize ten independent chains with ten different  values of $\sigma\in\{3.0, 3.5, \ldots, 7.5\}$. Following \cite{kou2006}, we set $\sigma$ to the value that leads to the best autocorrelation function among those that visit all  modes. This is 4.0 in case~(a) and $3.5$ in case~(b). The  acceptance rate is $0.048$ for case~(a) and $0.228$ for case~(b).

\begin{figure}[b!]
\begin{center}
\includegraphics[scale = 0.31]{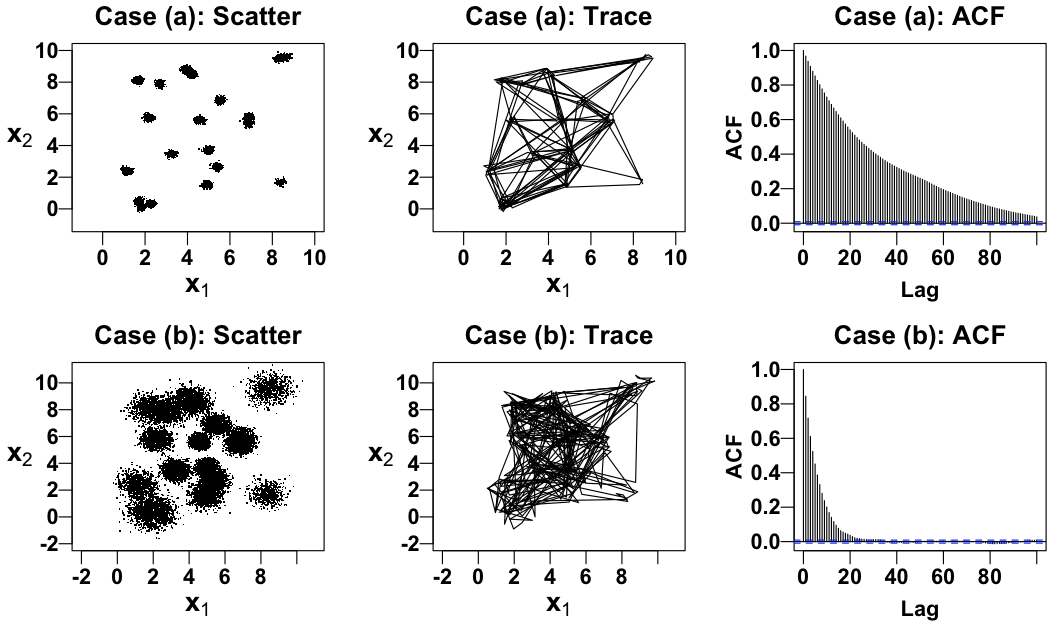}
\caption{Results of the RAM algorithm. The first column displays bivariate scatterplots for 50,000 samples, the middle column displays the bivariate trace plots for the last 2,000 samples for case (a) and  the last 1,000 samples for case (b), and the last column displays the autocorrelation functions for 50,000 samples of $x_1$. }
\label{example_summary2}
\end{center}
\end{figure}

Fig.~\ref{example_summary2} gives  bivariate scatterplots of the Monte Carlo sample of size 50,000 obtained with RAM for the two cases, bivariate trace plots of the last 2,000 iterations for case (a) and the last 1,000 iterations for case (b), and autocorrelation plots for $x_1$. Fig.~\ref{example_summary2} can be compared to Fig.~3 and Fig.~4 of \cite{kou2006},  which summarize the performance of EE and PT for cases (a) and (b), respectively.

To compare the accuracy of the moment estimates obtained with the  algorithms, we again follow \cite{kou2006} and run 20 independent chains using RAM. Table~\ref{summary_table} summarizes the comparisons, where the ratios of the mean squared error (MSE) of both EE and PT to that of RAM are all greater than one. The improvement is particularly striking for case~(b). These indicate that RAM leads to a more reliable proportion of iterations that are associated with each mode across the 20 runs.

\begin{table}[t!]
\caption{Moment estimates for cases (a) and (b) based on 20 independent chains, each of length 50,000, generated with RAM, EE (equi-energy sampler), and PT (parallel tempering). The results for EE and PT are from \cite{kou2006}, and presented in their original format: Sample average (sample standard deviation) over the 20 replications.}
{\footnotesize\begin{center}
\begin{tabular}{lrrrrccc}
\multirow{2}{*}{Case (a)} & \multicolumn{1}{c}{\multirow{2}{*}{Truth}}  & \multicolumn{1}{c}{\multirow{2}{*}{RAM}} & \multicolumn{1}{c}{\multirow{2}{*}{EE}} &   \multicolumn{1}{c}{\multirow{2}{*}{PT}} & MSE ratio & MSE ratio\\
&  & &  & & (EE/RAM) & (PT/RAM)\\
\hline
$E(x_1)$ &  $4.478$  & $4.4708$ $(0.091)$ & $4.5019$ $(0.107)$ & $4.4185$ $(0.170)$ & $1.44$ & $3.89$\\
$E(x_2)$ & $4.905$  & $4.9318$ $(0.101)$ & $4.9439$ $(0.139)$ & $4.8790$ $(0.283)$ & $1.91$ & $7.40$\\
$E(x^2_1)$ & $25.605$  & $25.5717$ $(0.900)$ & $25.9241$ $(1.098)$ & $24.9856$ $(1.713)$ & $1.61$ & $4.09$\\
$E(x^2_2)$ & $33.920$  & $33.2234$ $(1.100)$ & $34.4763$ $(1.373)$ & $33.5966$ $(2.867)$ & $1.69$ & $6.39$\\
\\
\multirow{2}{*}{Case (b)} & \multicolumn{1}{c}{\multirow{2}{*}{Truth}}  & \multicolumn{1}{c}{\multirow{2}{*}{RAM}} & \multicolumn{1}{c}{\multirow{2}{*}{EE}} &   \multicolumn{1}{c}{\multirow{2}{*}{PT}} & MSE ratio & MSE ratio\\
&  & &  & & (EE/RAM) & (PT/RAM)\\
\hline
$E(x_1)$ &  $4.688$  & $4.673$ $(0.026)$ & $4.699$ $(0.072)$ & $4.709$ $(0.116)$ & $5.89$ & $15.42$\\
$E(x_2)$ & $5.030$  & $5.029$ $(0.035)$ & $5.037$ $(0.086)$ & $5.001$ $(0.134)$ & $6.07$ & $15.33$\\
$E(x^2_1)$ & $25.558$  & $25.508$ $(0.263)$ & $25.693$ $(0.739)$ & $25.813$ $(1.122)$ & $7.87$ & $18.47$\\
$E(x^2_2)$ & $31.378$  & $31.456$ $(0.334)$ & $31.433$ $(0.839)$ & $31.105$ $(1.186)$ & $6.01$ & $12.59$\\
\end{tabular}
\end{center}}
\label{summary_table}
\end{table}

Finally, we compare the average evaluation cost of each algorithm by reporting  the expected total number of evaluations of the target density $\pi$  needed to obtain the final sample, including burn-in,  divided by the final sample size; we denote this quantity by $\textrm{N}_{\pi}^{\textrm{X}}$, where `X' specifies the algorithm. As detailed in Appendix~A,  $\textrm{N}_{\pi}^{\textrm{EE}}=16.0$ and $\textrm{N}_{\pi}^{\textrm{PT}}=5.8$. For RAM, $\textrm{N}_{\pi}^{\textrm{RAM}}=7.1$ in case (a) and $\textrm{N}_{\pi}^{\textrm{RAM}}=5.0$ in case (b)\footnote{The average number of proposals required by the forced downhill transition is 1.01 in case~(a) and 1.06 in case~(b), that of the uphill proposals is 4.70 in case~(a) and 2.57 in case~(b), and that of the downhill auxiliary variables is 1.39 in case~(a) and 1.35 in case~(b).}. More evaluations are needed for case~(a) because the area of near zero density  is much larger than that in case~(b), see Fig.~\ref{ee_contour}, and a forced uphill transition thus requires more proposals (and thus more evaluations). Nonetheless, the number of target density evaluations (and thus CPU time) required by RAM indicates that the gain of using RAM in terms of MSE is competitive.

\subsection{Example 2: High-dimensional  multimodal distributions}\label{ex_low_dim}
Consider an equal mixture of eight $d$-dimensional Gaussian distributions:
\begin{equation}
\pi(x)\propto\sum_{j=1}^8\exp\!\left(-\frac{1}{2}(x-\mu_j)^\top (x-\mu_j)\right)\!,\label{target_dist}
\end{equation}
where $x=(x_1, x_2, \ldots, x_d)^\top$  and the eight mean vectors are defined by setting their first three coordinates to the eight vertices of a cube of edge length ten situated with its corner at the origin and their remaining coordinates are filled with (10, 0) or (0, 10) repeatedly:
\begin{align}
\begin{aligned}\nonumber
\mu_1=(10, 10, 10, \textcolor{white}{1}0, 10, \textcolor{white}{1}0, 10, ..., \textcolor{white}{1}0, 10),~~\mu_2=(\textcolor{white}{1}0, \textcolor{white}{1}0, \textcolor{white}{1}0, 10, \textcolor{white}{1}0, 10, \textcolor{white}{1}0, ..., 10, \textcolor{white}{1}0),\\
\mu_3=(10, \textcolor{white}{1}0, 10, \textcolor{white}{1}0, 10, \textcolor{white}{1}0, 10, ..., \textcolor{white}{1}0, 10),~~\mu_4=(\textcolor{white}{1}0, 10, 10, \textcolor{white}{1}0, 10, \textcolor{white}{1}0, 10, ..., \textcolor{white}{1}0, 10),\\
\mu_5=(\textcolor{white}{1}0, \textcolor{white}{1}0, 10, \textcolor{white}{1}0, 10, \textcolor{white}{1}0, 10, ..., \textcolor{white}{1}0, 10),~~\mu_6=(\textcolor{white}{1}0, 10, \textcolor{white}{1}0, 10, \textcolor{white}{1}0, 10, \textcolor{white}{1}0, ..., 10, \textcolor{white}{1}0),\\
\mu_7=(10, \textcolor{white}{1}0, \textcolor{white}{1}0, 10, \textcolor{white}{1}0, 10, \textcolor{white}{1}0, ..., 10, \textcolor{white}{1}0),~~\mu_8=(10, 10, \textcolor{white}{1}0, 10, \textcolor{white}{1}0, 10, \textcolor{white}{1}0, ..., 10, \textcolor{white}{1}0).
\end{aligned}
\end{align}
Suppose that the first two modes, $\mu_1$ and $\mu_2$,  are known, perhaps from an initial search, while the other six modes are unknown. Here, we investigate RAM's ability to explore a high dimensional distribution by using it to sample \eqref{target_dist} with the five values of $d\in\{3, 5, 7, 9, 11\}$. We also compare RAM to both Metropolis and  PT, taking into account their average evaluation cost,  $\textrm{N}_{\pi}^{\textrm{X}}$, as defined in Section~\ref{ch3_example1}.

We set $q$ to be a $d$-dimensional Gaussian density with covariance matrix $\Sigma$. To achieve a reasonable acceptance rate, we first run two Metropolis chains each of length 5,000, initialized at the two known mode locations and using a Gaussian jumping rule with covariance matrix  $(2.38^2/d)\times I_d$, where $I_d$ is the identity matrix. We then set $\Sigma$ to the sample covariance matrix  of the combined sample from the two chains. To improve Metropolis' ability to jump between modes, we reset $\Sigma$ to the sample covariance matrix of the burn-in sample. This one-time adaptation  does not affect the validity of the resulting chain.

For each $d$, we run RAM ten times to obtain ten chains each of length 500,000, discarding the first 200,000 iterations of each chain as burn-in. RAM's average evaluation cost $\textrm{N}_{\pi}^{\textrm{RAM}}$ is 6.54 for $d=3$, 7.54 for $d=5$, 8.45 for $d=7$, 9.58 for $d=9$, and 10.77 for $d=11$. As $d$ increases, RAM requires more evaluations because it is more difficult to find a proposal that increases the density in the forced uphill transition.

For each $d$, we also obtain ten chains each using both Metropolis and PT with the same Gaussian jumping rule used by RAM. PT runs five parallel chains under five temperature levels, $2^{k}$ for $k=0, 1, \ldots, 4$, each of which uses Metropolis transitions. PT always proposes a single swap between a randomly chosen pair of chains under adjoining temperature levels at the end of each iteration. We determine the length of each chain and the burn-in size for Metropolis and PT by taking into account their average evaluation cost, denoted by $\textrm{N}^{\textrm{M}}_{\pi}$ and $\textrm{N}^{\textrm{PT}}_{\pi}$, respectively\footnote{With a caching strategy, PT evaluates the target once for a Metropolis transition under each of five temperature levels  and evaluates it twice  for a swap at the end of each iteration.}. For example, the length of each chain for Metropolis is 500,000$\times\textrm{N}^{\textrm{RAM}}_{\pi}/\textrm{N}^{\textrm{M}}_{\pi}$ and that for PT is 500,000$\times\textrm{N}^{\textrm{RAM}}_{\pi}/\textrm{N}^{\textrm{PT}}_{\pi}$ so that the (expected) total number of target density evaluations is the same for each algorithm.  We need to adjust the burn-in size by the average evaluation cost for a fair comparison because a large burn-in size improves the effectiveness of the one-time adaptation.

We use two numerical measures to evaluate each algorithm. The first  is the average number of the unknown modes that are discovered by each chain; we denote this by N$_{\textrm{dis}}$ $(\le6)$. The second  is the average frequency error rate \citep{kou2006}, denoted by $\textrm{F}_\textrm{err}=\sum_{i=1}^{10}\sum_{j=1}^8\vert \textrm{F}_{i, j}-1/8\vert/80$, where $\textrm{F}_{i, j}$ is the proportion of iterations in chain~$i$ whose nearest mode measured by the Euclidean distance is $\mu_j$.

\begin{table}[t!]
{\small \caption{The sampling results include the length of each chain before discarding burn-in; the number of burn-in iterations; N$_{\pi}$ = the average number of target density evaluations at each iteration; $\textrm{N}_d$ = the average number of downhill proposals for RAM; $\textrm{N}_u$ = the average number of uphill proposals for RAM; {$\textrm{N}_z$ =} the average number of downhill proposals for the auxiliary variable for RAM; acceptance rate; N$_{\textrm{dis}}$ = the average number of the unknown modes that are discovered by each chain; and F$_{\textrm{err}}= \sum_{i=1}^{10}\sum_{j=1}^8\vert \textrm{F}_{i, j}-1/8\vert/80$, where $\textrm{F}_{i, j}$ is the proportion of iterations in chain~$i$ whose nearest mode is $\mu_j$.}
\label{table_high}
\begin{center}
\begin{tabular}{cccccccr}
\multirow{2}{*}{$d$}&\multirow{2}{*}{Kernel}& Length of a chain&  $\textrm{N}_{\pi}$  &  Acceptance & \multirow{2}{*}{N$_{\textrm{dis}}$} & \multirow{2}{*}{F$_{\textrm{err}}$}\\
&& (burn-in size)  & ($\textrm{N}_d$, $\textrm{N}_u$, $\textrm{N}_z$)& rate& \\
\hline
\multirow{3}{*}{$3$}&Metropolis& 3,272,000 (1,308,800)  & 1  & 0.036 & 6.0 & 0.021\\%0.9347
&PT& \textcolor{white}{2,}467,429 \textcolor{white}{2,}(186,971) &  7 &  0.025 & 6.0 & 0.025 \\
&RAM& \textcolor{white}{2,}500,000 \textcolor{white}{2,}(200,000) & 6.544 & 0.101 & 6.0 & 0.019 \\
&& & (1.014, 4.210, 1.320) &  &  &  \\\\
\multirow{3}{*}{$5$}&Metropolis& 3,768,500 (1,507,400)  & 1  & 0.019 & 6.0 & 0.047\\%0.9347
&PT& \textcolor{white}{2,}538,357 \textcolor{white}{2,}(215,343) &  7 & 0.041  & 6.0 & 0.041\\
&RAM& \textcolor{white}{2,}500,000 \textcolor{white}{2,}(200,000) & 7.537 & 0.052 & 6.0 & 0.038 \\
& &  &  (1.009, 5.222, 1.306) & & & \\\\
\multirow{3}{*}{$7$}&Metropolis& 4,220,500 (1,688,200)  & 1  & 0.014 & 5.8 & 0.209\\%0.9347
&PT& \textcolor{white}{2,}602,929 \textcolor{white}{2,}(241,171) &  7 &  0.058 & 6.0 & 0.058\\
&RAM& \textcolor{white}{2,}500,000 \textcolor{white}{2,}(200,000) & 8.441 & 0.036 & 6.0 & 0.075 \\
& &  & (1.006, 6.136, 1.299) & &  & \\\\
\multirow{3}{*}{$9$}&Metropolis& 4,734,000 (1,893,600)  & 1  &  0.012 & 5.6 & 0.312\\%0.9347
&PT& \textcolor{white}{2,}676,286 \textcolor{white}{2,}(270,514) &  7 & 0.075  & 6.0 & 0.075\\
&RAM& \textcolor{white}{2,}500,000 \textcolor{white}{2,}(200,000) & 9.468  & 0.029 & 5.7 & 0.182 \\
& & & (1.005, 7.171, 1.292) &  & & \\\\
\multirow{3}{*}{$11$}&Metropolis& 5,350,000 (2,140,000)  & 1  & 0.023 & 5.3 & 0.512\\%0.9347
&PT& \textcolor{white}{2,}764,286 \textcolor{white}{2,}(305,714) &  7 &  0.004 & 6.0 & 0.108\\
&RAM& \textcolor{white}{2,}500,000 \textcolor{white}{2,}(200,000) & 10.700 & 0.021 & 5.5 & 0.267 \\
& & & (1.003, 8.416, 1.281) & & & \\
\end{tabular}
\end{center}}
\end{table}

Table \ref{table_high} summarizes the results, and shows that using the same jumping rule, RAM is never worse than Metropolis in terms of $\textrm{N}_{\textrm{dis}}$ and  $\textrm{F}_{\textrm{err}}$ regardless of  dimension, and the improvement on $\textrm{F}_{\textrm{err}}$ can be substantial. It also shows that RAM's $\textrm{F}_{\textrm{err}}$ starts off smaller than that of PT but deteriorates faster than PT's once $d>5$, and that PT discovers all six modes for every $d$. This demonstrates the value of fine tuning particularly in higher dimensions for PT, including the number of parallel chains, temperature and proposal scale at each chain, and the number and rate of swaps at each iteration. Therefore, if one can afford the tuning cost, then PT has much to recommend it, especially in high dimensions.

 \subsection{Example 3: Sensor network localization}\label{example4}
For high dimensional sampling, a blocked Gibbs sampler \citep{geman1984stochastic} is sometimes more convenient and intuitive than direct Metropolis sampling. Here, we consider a realistic example from \cite{ihler2005nonparametric}: Searching for unknown sensor locations within a network using the noisy distance data. This is called sensor network localization \citep{ihler2005nonparametric, lan2014wormhole}. This problem is known to produce a high-dimensional, banana-shaped, and multimodal joint posterior distribution.

\begin{figure}[b!]
\begin{center}
\includegraphics[scale = 0.4]{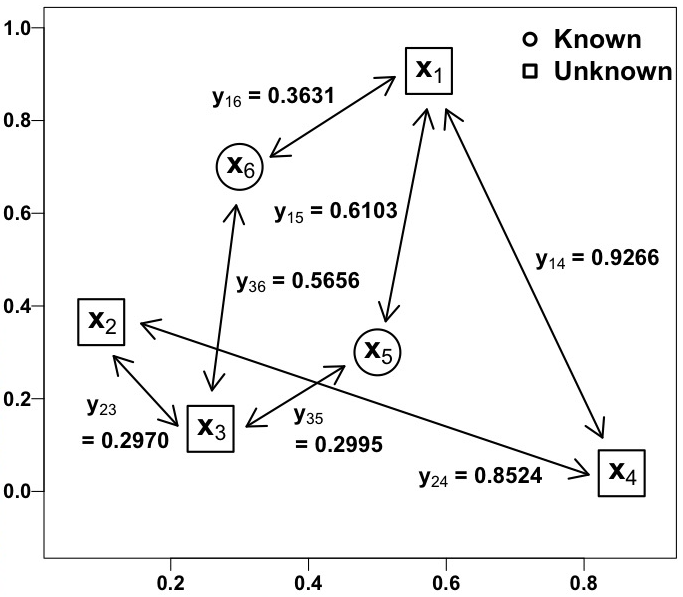}
\caption{The simulated distances $y_{ij}~(=y_{ji})$ among the six stationary sensor locations, $x_1, x_2, \ldots, x_6$, are displayed if observed. The observation indicator $w_{ij}~(=w_{ji})$ is one if $y_{ij}$ is specified and is zero otherwise. The  locations of the sensors are $x_{1}=(0.57, 0.91)$, $x_{2}=(0.10, 0.37)$, $x_{3}=(0.26, 0.14)$, $x_{4}=(0.85, 0.04)$, $x_{5}=(0.50, 0.30)$, and $x_{6}=(0.30, 0.70)$, where the first four locations, $x_1$, $x_2$, $x_3$, and $x_4$, are assumed to be unknown. }
\label{table_sensor_data}
\end{center}
\end{figure}

Modifying \cite{lan2014wormhole}'s simulation setting\footnote{We remove some locations and adjust observed distances to make a simpler model, yet the  resulting posterior distributions have more complicated shapes.}, we suppose there are six stationary sensors scattered on a two dimensional space, and let $x_k^\top=(x_{k1}, x_{k2})$ denote the two dimensional coordinates of the location of sensor~$k$  for $k=1, 2, \ldots, 6$. We assume that the locations of the last two sensors, $x_5$ and $x_6$, are known and the locations of the other sensors, $x_1, x_2, x_3$, and $x_4$, are unknown parameters of interest. The Euclidean distance between two sensors, $x_i$ and $x_j$, denoted by $y_{ij}~(=y_{ji})$, is observed with a distance-dependent probability and Gaussian measurement error  for $i=1, 2, \ldots, 5$ and $j=i+1, \ldots, 6$. The probability distributions for the observed data are
\begin{equation}\nonumber%\label{sensor_data_prob}
w_{ij}\mid x_1, \ldots, x_4~ \sim \textrm{Bernoulli}\!\left(\exp\!\left(-\frac{\Vert x_i - x_j \Vert^2}{2\times0.3^2}\right)\right)\!\\
\end{equation}
and
\begin{equation}\nonumber%\label{sensor_data_prob}
y_{ij}\mid (w_{ij} = 1), x_1, \ldots, x_4~ \sim \textrm{N}_{1}\!\left(\Vert x_i - x_j\Vert,~ 0.02^2\right)\!,
\end{equation}
where $w_{ij}~(=w_{ji}$) is an  indicator variable that equals one if the distance between $x_i$ and $x_j$ is observed. Simulated distances $y_{ij}$ are displayed in Fig.~\ref{table_sensor_data} where $w_{ij}=1$ if $y_{ij}$ is specified and zero otherwise. For each unknown location, we assume a diffuse bivariate Gaussian prior distribution with mean (0, 0) and covariance matrix $10^2\times I_2$. The eight dimensional likelihood function is thus
\begin{align}
\begin{aligned}\nonumber
L(x_1, x_2, x_3, x_4)\propto &\prod_{j>i} \bigg[\exp\!\left(-\frac{(y_{ij}-\Vert x_i - x_j\Vert)^2}{2\times 0.02^2} \right)\\
&\times \exp\!\left(-\frac{w_{ij}\times \Vert x_i - x_j \Vert^2}{2\times0.3^2}\right)\times\left(1-\exp\!\left(-\frac{ \Vert x_i - x_j \Vert^2}{2\times0.3^2}\right)\right)^{1-w_{ij}}\bigg]
\end{aligned}
\end{align}
and the full posterior distribution is
\begin{equation}\label{full_sensor}
\pi(x_1, x_2, x_3, x_4 \mid y, w) \propto L(x_1, x_2, x_3, x_4)\times \exp\left(-\frac{\sum_{k=1}^4x_k^\top x_k}{2 \times 10^2}\right),
 \end{equation}
where $y=\{y_{ij}, i>j\}$ and $w=\{w_{ij}, i>j\}$. This model may suffer from non-identifiability when the number of observed distances is small because  unknown locations appear in the likelihood only through distances; if $y_{ij}$ is observed between an unknown $x_i$ and a known $x_j$,  the posterior distribution of $x_i$ may form a circle around $x_j$ without further observations.

We sample \eqref{full_sensor} using a Gibbs sampler by iteratively sampling the  four bivariate conditionals denoted by $
\pi_i(x_i\mid x_j, j\not =i, y, w)$ for $i=1, 2, 3, 4$.
Since none of these is a standard distribution, we use Metropolis, RAM, or tempered transition (TT) \citep{neal1996} kernels that are invariant with respect to each conditional distribution; see Appendix~B for details of TT, jumping rules, and initial values. To sample $x_k$ from a RAM kernel that is marginally invariant to $\pi_k$, we must keep track of the auxiliary variable during the run, i.e., \{$z^{(i)}_k$, $i=0, 1, 2, \ldots$\}. At iteration~$i$, we  sequentially draw {$x_k'\sim q^{\textrm{D}}(x_k'\mid x_k^{(i-1)})$}, {$x_k^\ast\sim q^{\textrm{U}}(x_k^\ast\mid x_k')$}, and {$z_k^\ast\sim q^{\textrm{D}}(z_k^\ast\mid x_k^\ast)$}. We set $(x_k^{(i)}, z_k^{(i)})$ to $(x_k^\ast, z_k^\ast)$ with  probability $\alpha^\textrm{J}(x_k^\ast, z_k^\ast\mid  x_k^{(i-1)}, z_k^{(i-1)})$ given  in \eqref{joint_accept_prob2}, and set $(x_k^{(i)}, z_k^{(i)})$ to $(x_k^{(i-1)}, z_k^{(i-1)})$ otherwise.  Because \{$z^{(i)}_k$, $i=0, 1, 2, \ldots$\} are introduced solely to enable sampling $x_k$ from a RAM kernel, only $x_k^{(i)}$ is used to sample the other locations, and $z_k^{(i)}$ is  used to draw $x_k^{(i+1)}$ at the next iteration.

For a fair comparison, we set the length of each chain to have the same average number of evaluations of $\pi_i$'s per iteration. As before, we use $\textrm{N}_{\pi}^{\textrm{X}}$ to denote the average evaluation cost. We first implement RAM within a Gibbs sampler  for 220,000 iterations with the first 20,000 as burn-in, resulting in $\textrm{N}_{\pi}^{\textrm{RAM}}=36.13$, i.e., about nine density evaluations are required to sample each of the $\pi_i$'s (with caching). Since $\textrm{N}_{\pi}^{\textrm{M}}=4$ and $\textrm{N}_{\pi}^{\textrm{TT}}=24$ (with caching), we set the length of each Metropolis chain and TT chain respectively to 220,000$\times\textrm{N}_{\pi}^{\textrm{RAM}}/\textrm{N}_{\pi}^{\textrm{M}}$ and 220,000$\times\textrm{N}_{\pi}^{\textrm{RAM}}/\textrm{N}_{\pi}^{\textrm{TT}}$. However, unlike the previous example where there is a one-time adaption and hence it is important to adjust for the burn-in length as well, here we discard the first 20,000 iterations as burn-in for all three algorithms. This burn-in size is sufficient to remove the effect of random initial values of the algorithms.

\begin{table}[b!]
{\small\caption{The sampling results summarize the length of each chain (including the 20,000 burn-in iterations); $\textrm{N}_{\pi}=$ the average number of evaluating $\pi_1$, $\pi_2$, $\pi_3$, and $\pi_4$ at each iteration; details of $\textrm{N}_{\pi}$ for each location; and the acceptance rates.}
\label{table_sensor2}
\begin{center}
\begin{tabular}{ccccc}
\multirow{2}{*}{Kernel}& \multirow{2}{*}{Length of a chain} & \multirow{2}{*}{$\textrm{N}_{\pi}$}  &  Details of $\textrm{N}_{\pi}$ &  Acceptance\\
&  & &($\textrm{N}_d$, $\textrm{N}_u$, $\textrm{N}_z$)& rate\\
\hline
\multirow{4}{*}{Metropolis}& \multirow{4}{*}{1,987,150}  & \multirow{4}{*}{$\textrm{N}_{\pi}^{\textrm{M}}=4$} & \multirow{4}{*}{1 for each of $x_1, \ldots, x_4$}  & 0.00057  for $x_1$\\%0.9347
&&& & 0.00151  for $x_2$\\%0.9347
 &&&  & 0.00053 for $x_3$\\%0.9347
 &&&  & 0.00115 for $x_4$\\\\%0.9347
\multirow{3}{*}{Tempered}& \multirow{4}{*}{331,192}  & \multirow{4}{*}{$\textrm{N}_{\pi}^{\textrm{TT}}=24$} & \multirow{4}{*}{6 for each of $x_1, \ldots, x_4$}  & 0.00360  for $x_1$\\%0.9347
\multirow{3}{*}{transitions}&&  &  & 0.01034  for $x_2$\\%0.9347
 && & & 0.00369  for $x_3$\\%0.9347
 && & & 0.00918  for $x_4$\\\\%0.9347
\multirow{4}{*}{RAM}& \multirow{4}{*}{220,000}  &  \multirow{4}{*}{$\textrm{N}_{\pi}^{\textrm{RAM}}=36.13$} & 9.40 for $x_1$ (1, 7.33, 1.07)  & 0.00349  for $x_1$\\%0.9347
&& & 8.64 for $x_2$ (1, 6.56, 1.08) & 0.00830  for $x_2$\\%0.9347
 && & 9.22 for $x_3$ (1, 7.16, 1.06) & 0.00353  for $x_3$\\%0.9347
 && & 8.87 for $x_4$ (1, 6.74, 1.13) & 0.00730  for $x_4$\\%0.9347
\end{tabular}
\end{center}}
\end{table}

Table~\ref{table_sensor2} summarizes the configurations of the samplers and their acceptance rates. RAM improves the acceptance rate of Metropolis by a factor at least of 5.5 given the same jumping rule without additional tuning. TT improves the acceptance rates even further by a factor of at least 6.3 (relative to Metropolis), but it requires additional tuning of the number of temperature levels, temperature, and jumping scale at each temperature level.

\begin{figure}[t!]
\begin{center}
\includegraphics[scale = 0.455]{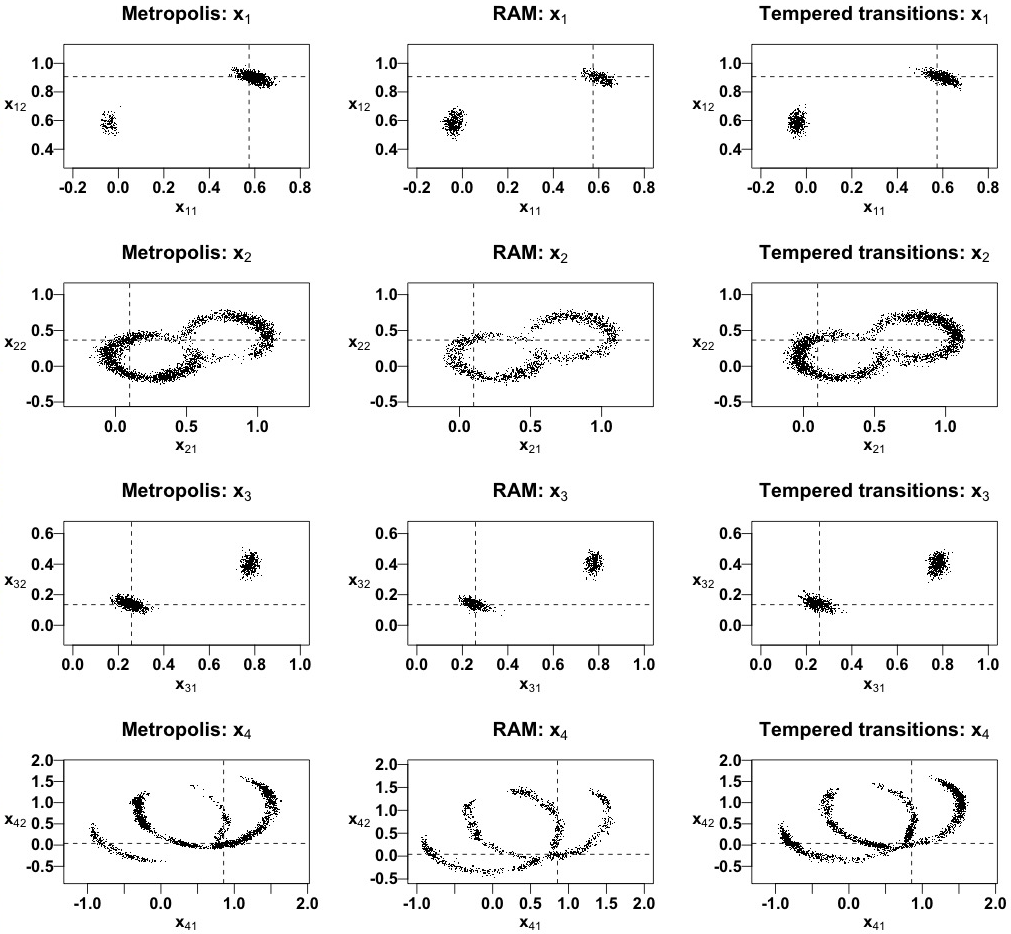}
\caption{Scatterplots of the posterior sample of each location (rows) obtained by different samplers (column). The coordinates of the unknown sensors are denoted by dashed lines.}
\label{res_sensor}
\end{center}
\end{figure}

Fig.~\ref{res_sensor} gives scatterplots of the posterior samples of each unknown sensor location (rows) obtained by the three samplers (columns), where the dashed lines indicate the coordinates of the true location. The RAM sample is more dispersed than that of Metropolis, especially for $x_1$, $x_2$, and $x_4$, with the same jumping rule, and is as dispersed as that of TT without subtle tuning that TT requires. The posterior samples of both Metropolis and TT, however, are denser than that of RAM because of their larger sample sizes.

\begin{figure}[t!]
\begin{center}
\includegraphics[scale = 0.47]{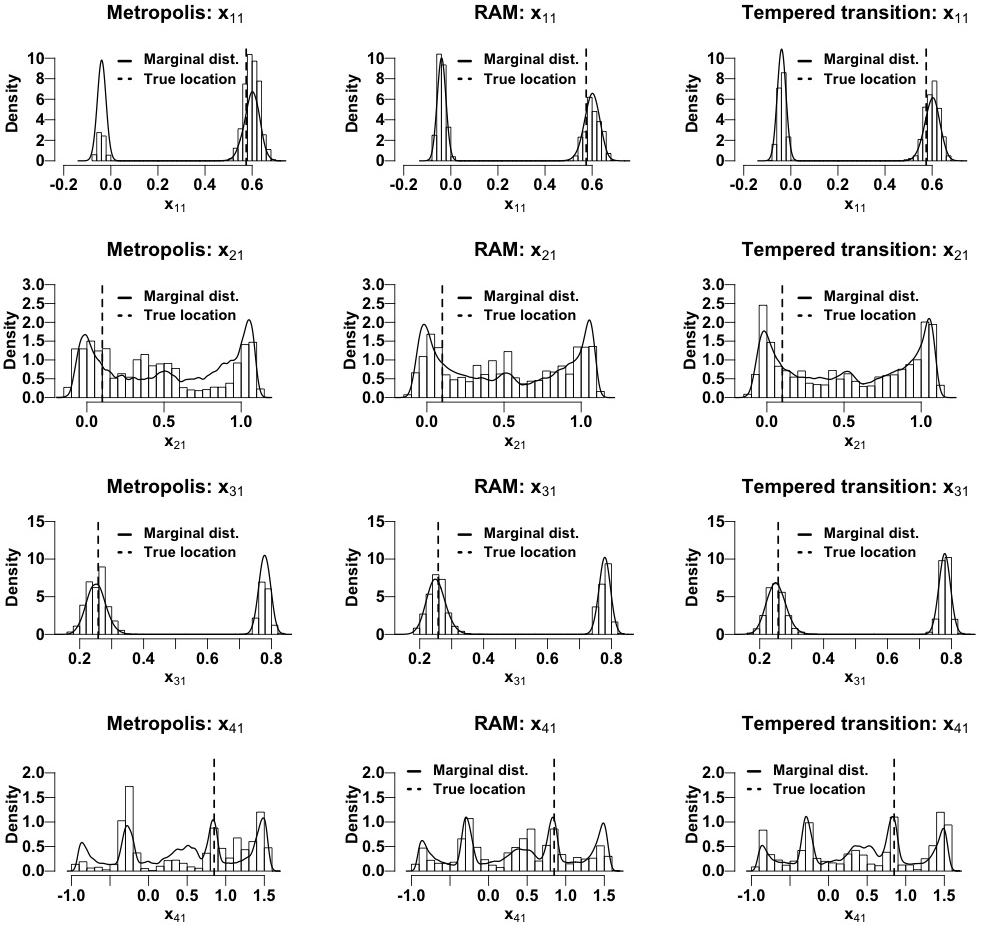}
\caption{Histograms of the posterior sample of each first coordinate (rows) obtained by different kernels (columns). In each histogram, the marginal posterior density based on twenty million posterior samples obtained with each sampler is superimposed. The vertical dashed lines indicate the true sensor locations. }
\label{res_sensor2}
\end{center}
\end{figure}

Fig.~\ref{res_sensor2} compares the relative sizes of modes for the first coordinate of each unknown location (rows) obtained by each sampler (columns).  In each histogram, we superimpose the marginal posterior density based on twenty million posterior draws obtained from each sampler after confirming that the shapes of the posterior densities obtained in this manner are almost identical for the three algorithms. The vertical dashed lines indicate the true sensor locations. RAM represents all four distributions better than Metropolis does, and it does as well as TT, but without the tuning requirement of the latter.

 \subsection{Example 4: Strong lens time delay estimation}\label{example2}

Our final numerical illustration targets a multimodal distribution where one mode is extremely distant from the others. This multimodal distribution arises from the applied astrophysical problem that originally motivated the development of  RAM; see \cite{tak2016c} for details.  Here we  review the problem and discuss a new efficient algorithm.

When there is a massive galaxy between a highly luminous quasar and the Earth, the gravitational field of the  galaxy may act as a  lens, bending the light  emitted by the quasar. This may produce two (or more) slightly offset images of the quasar, an effect known as strong gravitational lensing \citep{schneider2006}.  There may be a time delay between the images in that their light follows different paths with different travel times. Thus, temporal features in time series of the brightness of each image appear shifted in time. The time delay is, for example, used to calculate the current expansion rate of the Universe, i.e., the Hubble constant \citep{refsdal1964}.

\begin{figure}[b!]
\begin{center}
\includegraphics[scale = 0.43]{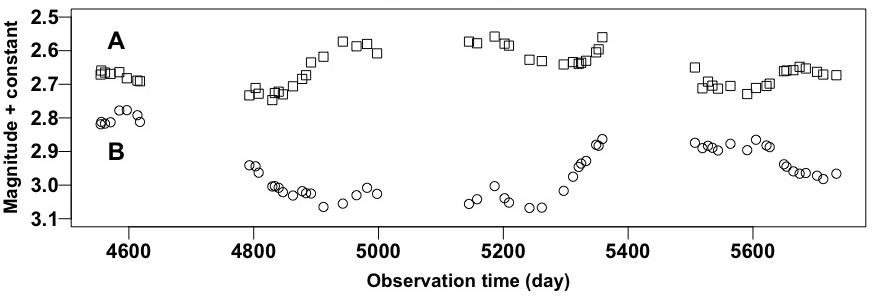}
\caption{Two observed time series of doubly-lensed quasar Q0957+561 \citep{hainline2012new}.  Time series $A$ is denoted by squares and time series $B$ is denoted by circles. Magnitude is an astronomical  measure of brightness. Both time series are plotted with an offset (constant) in magnitude, but this  does not affect the time delay estimation.}
\label{tdc1}
\end{center}
\end{figure}

Fig.~\ref{tdc1} displays two irregularly-observed  time series of the brightness of the doubly-lensed quasar Q0957+561 \citep{hainline2012new}; the two time series are labeled $A$ and $B$. Brightness is reported on a magnitude scale where smaller values correspond to brighter images. Let $x\equiv \{x_{1}, \ldots, x_n\}$ and $y\equiv\{y_1, \ldots, y_n\}$ denote the $n$  magnitudes irregularly observed at time $t\equiv\{t_1, \ldots, t_n\}$ in time series $A$ and $B$, respectively. Let $\delta\equiv \{\delta_1,  \ldots, \delta_n\}$ and $\eta\equiv\{\eta_1, \ldots, \eta_n\}$ represent the  $n$ known measurement standard deviations for $x$ and $y$, respectively. There are fifty seven observations in each time series, i.e.,  $n=57$.

We assume that for each observed time series there is an unobserved underlying brightness curve. Let $X(t)\equiv \{X(t_1), \ldots, X(t_n)\}$ denote the latent magnitudes for time series~$A$ and $Y(t)\equiv \{Y(t_1), \ldots, Y(t_n)\}$ denote those for time series~$B$. We further assume that one of the latent brightness curves is a shifted version of the other, i.e., 
\begin{equation}\label{fundamental}
Y(t_j)=X(t_j-\Delta)+\beta_0,
\end{equation}
where $\Delta$ is the unknown time delay and $\beta_0$ is an unknown magnitude offset.

The observed magnitudes given the latent magnitudes are assumed to be independent Gaussian variables:
\begin{equation}
x_j\mid X(t_j) \sim \textrm{N}_1\!\left(X(t_j),~  \delta^2_j\right)\!\quad \textrm{and} \quad
y_j\mid Y(t_j ) \sim \textrm{N}_1\!\left(Y(t_j),~ \eta^2_j\right)\!.\label{lik2}
\end{equation}
Using \eqref{fundamental}, we can express the model for $y$ in \eqref{lik2} as
\begin{equation}\label{lik3}
y_j\mid X(t_j - \Delta), \Delta, \beta_0 \sim \textrm{N}_1\!\left(X(t_j - \Delta) + \beta_0,~ \eta^2_j\right)\!.
\end{equation}

We assume $X(\cdot)$ follows an Ornstein-Uhlenbeck process \citep{kelly2009}. Solving the resulting  stochastic differential equation  yields the sampling distribution of $X(t^\Delta)$, where $t^\Delta\equiv(t^\Delta_1, \ldots, t^\Delta_{2n})^\top$ contains the sorted $2n$ times among the $n$ observation times, $t$, and the $n$ time-delay-shifted observation times, $t-\Delta$. Specifically,
\begin{align} 
\begin{aligned}\label{ou_prior}
X(t_1^\Delta)\mid \Delta, \theta &~\sim~ \textrm{N}_1\!\left(\mu,~ \frac{\tau\phi^2}{2}\right)\!, ~~~\textrm{and for}~ j=2, 3, \ldots, 2n,\\
~~X(t_j^\Delta)\mid X(t_{j-1}^\Delta), \Delta, \theta &~\sim~ \textrm{N}_1\!\left(\mu + a_j\big(X(t_{j-1}^\Delta) - \mu\big) ,~\frac{\tau \phi^2}{2}(1-a^2_j)\right)\!,
\end{aligned}
\end{align}
where $\theta\equiv (\mu, \phi^2, \tau)^\top$ and $a_j=\exp( -(t^\Delta_j-t^\Delta_{j-1})/\tau)$.

Following \cite{tak2016c}, we set independent priors for the model parameters:
\begin{align}
\begin{aligned}\label{priors}
\Delta &\sim \textrm{Uniform}[-1178.939,~1178.939],~\beta_0\sim \textrm{Uniform}[-60, 60],\\
\mu&\sim\textrm{Uniform}[-30, 30], ~\phi^2\sim \textrm{inverse-Gamma}(1, 2/10^{7}),~ \tau\sim \textrm{inverse-Gamma}(1, 1).
\end{aligned}
\end{align}
The resulting joint posterior density function is
\begin{align}
\begin{aligned}\label{target_post}
\pi(X(t^\Delta), \Delta, \beta_0, \theta \mid x, y)\propto&\left[\prod_{j=1}^{n} f_1\!\left(x_{j}\mid X(t_j)\right) \times f_2\!\left(y_{j}\mid X(t_j -\Delta), \Delta, \boldsymbol{\beta}\right)\right]\\
\times~ g(X(t^\Delta_1)\mid \Delta, \boldsymbol{\theta})&\times\left[\prod_{j=2}^{2n} g(X(t^\Delta_j)\mid X(t^\Delta_{j-1}), \Delta, \boldsymbol{\theta})\right]\times h( \Delta, \beta_0, \theta),
\end{aligned}
\end{align}
where the density functions, $f_1$, $f_2$, $g$, and $h$ are defined by \eqref{lik2}--\eqref{priors}, respectively.

\begin{table}[b!]
\caption{A Metropolis-Hastings within Gibbs sampler for the time delay model. We draw $\Delta$ from a kernel that is invariant to $\pi_{11}$ and draw  $X(t^\Delta)$ from $\pi_{12}$ if $\Delta$ is newly updated.}\label{al2}
\bigskip
\begin{tabular}{l}
\hline
Set initial values $\Delta^{(0)}$,  $X^{(0)}(t^{\Delta^{(0)}})$, $\beta_0^{(0)}$, and $\theta^{(0)}$. For $i=1, 2, \ldots$,\\
\emph{Step} 1: Draw $\left(X^{(i)}(t^{\Delta^{(i)}}), \Delta^{(i)}\right)\sim  \pi_1\left(X(t^\Delta), \Delta \mid \beta^{(i-1)}_0, \theta^{(i-1)}\right)$\\
$~~~~~~~~~~~~~~~~~~~~~~~~~~~~~~~~~~~~~~~~~~$=~$\pi_{11}\left(\Delta \mid \beta^{(i-1)}_0, \theta^{(i-1)}\right) \pi_{12}\left(X(t^\Delta)\mid \Delta, \beta^{(i-1)}_0, \theta^{(i-1)}\right).$\\
 \emph{Step} 2: Draw $\beta^{(i)}_0\sim \pi_2\left(\beta_0 \mid \theta^{(i-1)}, X^{(i)}(t^{\Delta^{(i)}}), \Delta^{(i)}\right)$.\\
\emph{Step} 3: Draw $\theta^{(i)} \sim \pi_3\left(\theta \mid X^{(i)}(t^{\Delta^{(i)}}), \Delta^{(i)}, \beta^{(i)}_0\right).$
\end{tabular}
\end{table}

To sample from \eqref{target_post},  we adopt an MH within Gibbs sampler \citep{tierney1994} composed of the three steps  shown in Table~\ref{al2}; see Appendices A--C of \cite{tak2016c} for details. We suppress conditioning on $x$ and $y$  here and elsewhere. Because we cannot directly sample $\pi_{11}(\Delta\mid \beta_0, \theta)$ in \emph{Step}~1 and the marginal posterior distribution of $\Delta$  is often multimodal,  we draw $\Delta$ using one of four algorithms: (i) Metropolis, (ii) Metropolis with a mixture jumping
rule, (iii)~RAM, or (iv) TT. The mixture jumping rule   generates a proposal from the Gaussian jumping rule used by Metropolis with probability 0.5 and from the  prior distribution of~$\Delta$ otherwise. To sample $\Delta$ using the RAM kernel,  we additionally keep track of the auxiliary variable during the run, i.e., \{$z^{(i)}$, $i=0, 1, 2, \ldots$\}. At iteration~$i$, we draw {$\Delta'\sim q^{\textrm{D}}(\Delta'\mid \Delta^{(i-1)})$}, {$\Delta^\ast\sim q^{\textrm{U}}(\Delta^\ast\mid \Delta')$}, and {$z^\ast\sim q^{\textrm{D}}(z^\ast\mid \Delta^\ast)$} sequentially. We set $(\Delta^{(i)}, z^{(i)})$ to $(\Delta^\ast, z^\ast)$ with  probability $\alpha^\textrm{J}(\Delta^\ast, z^\ast\mid  \Delta^{(i-1)}, z^{(i-1)})$ given  in \eqref{joint_accept_prob2}, and set $(\Delta^{(i)}, z^{(i)})$ to $(\Delta^{(i-1)}, z^{(i-1)})$ otherwise.  Because \{$z^{(i)}$, $i=0, 1, 2, \ldots$\} are introduced solely to enable sampling $\Delta$ from the RAM kernel, only $\Delta^{(i)}$ is used to sample $X(t^\Delta)$, $\beta_0$, and $\theta$  in the other steps in  Table~\ref{al2}, and $z^{(i)}$ is used to draw $\Delta^{(i+1)}$ at the next iteration.
% See Appendix~D for more implementation details.

Specifically, we  fit the time delay model using the MH within Gibbs sampler equipped with TT  for $\Delta$ first, initiating a single long chain of length 5,050,000 at the center of the entire range of $\Delta$, i.e., $\Delta^{(0)}=0$. We set the initial values of the other parameters as follows; $\beta_0^{(0)}=\sum_{j=1}^n\{y_j-x_j\}/n=-0.113$, $\mu^{(0)}=\sum_{j=1}^nx_j/n={2.658}$, $\phi^{(0)}=0.01$,  $\tau^{(0)}=200$, and $X^{(0)}(t^{\Delta^{(0)}})$ is a vector of $x$ and $y-\beta_0^{(0)}$ that are sorted in time,  $t$ for $x$ and $t-\Delta$ for $y-\beta_0^{(0)}$.   Multiple initial values spread across the entire range result in nearly identical posterior distributions.  We discard the first 50,000 draws as  burn-in.   For the tuning parameters of TT, we set five temperature levels, $T_j=4^j$ for $j=1, \ldots, 5$, and corresponding  jumping scales for  Metropolis updates, $\sigma_j=500 \times 1.2^{j-1}$, so that $\sigma_5$ (= 1,037) is about a half of the length of the range of $\Delta$. Using the same initial values ($z^{(0)}=\Delta^{(0)}$ for RAM), we obtain an additional chain using each of the MH within Gibbs sampler equipped with Metropolis, RAM, and Metropolis with a mixture jumping rule. In all these cases, we set  $q$ to be Gaussian with $\sigma=700$, i.e., about one third length of the entire range and similar to the jumping scale of TT at the middle temperature level ($\sigma_3=720$). This value of $\sigma$ should be advantageous to Metropolis because it roughly equals the distance between the modes. Since Metropolis, RAM, and Metropolis with a mixture jumping rule take less CPU time  than  TT, we run longer chains of the three algorithms to match the CPU time, discarding the  first 50,000 iterations of each as burn-in; see Appendix~C for details of the average number of $\pi_{11}$ evaluations.

\begin{table}[b!]
{\small \caption{The length of a chain including burn-in; acceptance rate for $\Delta$;  and $\textrm{N}_{\textrm{jumps}}$ = the total number of jumps  between the two distant modes during the post burn-in run. }
\label{table_tdc}
\begin{center}
\begin{tabular}{lccr}
  &  Length of &   Acceptance  &  \multirow{2}{*}{$\textrm{N}_{\textrm{jumps}}$}\\
  &  a chain   &  rate & \\
\hline
\textcolor{white}{i}(i)\textcolor{white}{i} Metropolis & \multicolumn{1}{r}{22,266,816}  & 0.0150 & 144\\
\textcolor{white}{i}\!(ii)\textcolor{white}{i}\! Metropolis with mixture jumping rule & \multicolumn{1}{r}{19,710,188}  & 0.0129 & 190\\
(iii) RAM & \multicolumn{1}{r}{7,111,612} &  0.0508 & 326 \\
(iv) Tempered transitions &  \multicolumn{1}{r}{5,050,000}  & 0.3022 & 311\\
%RAM ($\sigma=500$) & \multicolumn{1}{r}{7,111,612} &  0.0498 & 695 \\
\end{tabular}
\end{center}}
\end{table}

Table \ref{table_tdc} summarizes the results from running each algorithm for nearly the same CPU time (28,352 seconds).  Overall, given the same jumping rule and without additional tuning,  RAM improves upon both versions of Metropolis; the total number of jumps between the two distant modes in the post burn-in sample, denoted by $\textrm{N}_{\textrm{jumps}}$,  is at least 1.7 times higher for RAM,  and RAM's acceptance rate is at least 3.4 times higher. With additional tuning of the number of rungs, temperature, and jumping scale, TT performs no better than RAM in terms of $\textrm{N}_{\textrm{jumps}}$ but its acceptance rate  is about 5.9 times higher than Metropolis.

\begin{figure}[b!]
\begin{center}
\includegraphics[scale = 0.395]{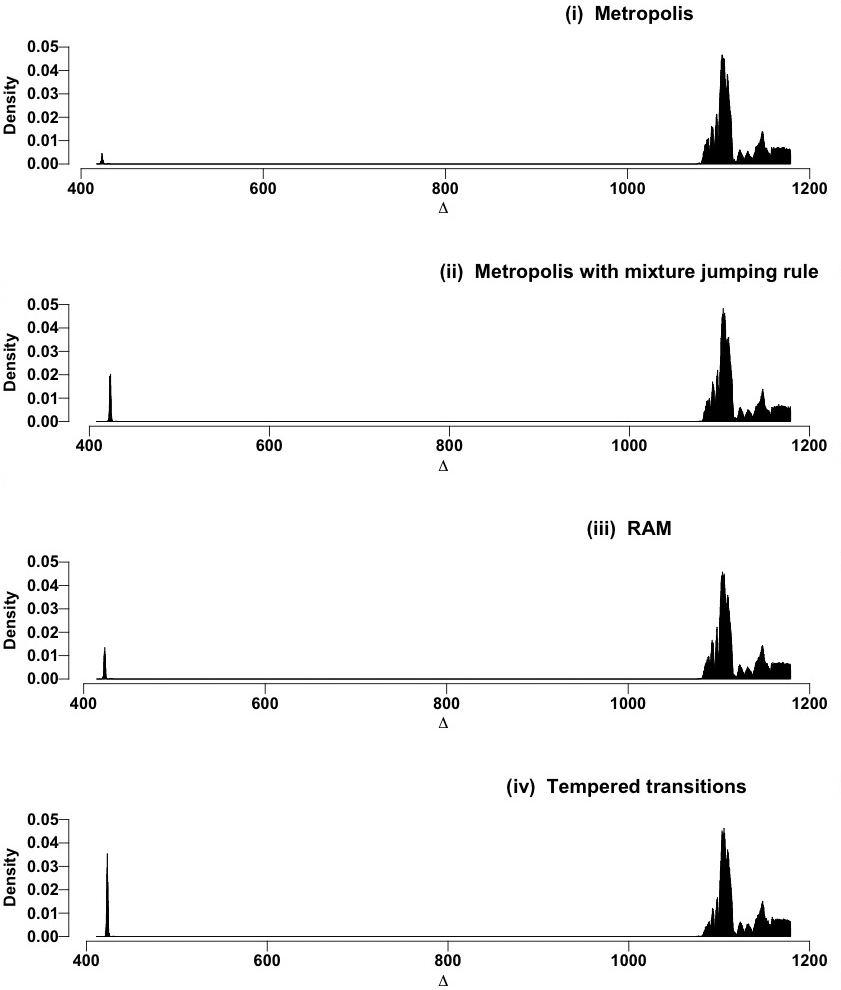}\includegraphics[scale = 0.394]{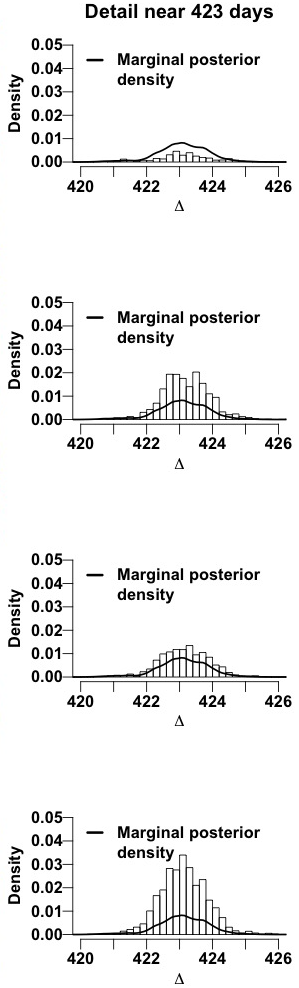}
\caption{Results of running the algorithms for nearly the same CPU time. The rows represent the four samplers. The first column displays the histograms based on the posterior sample of $\Delta$ and the second column focuses on the mode near 423 days. In the second column, we superimpose the  posterior density of $\Delta$ obtained by an oracle sampler  (assuming  the mode locations are known) to check the reliability of the relative sizes of the modes.}
\label{tdc2}
\end{center}
\end{figure}

The first column of Fig.~\ref{tdc2} displays histograms of the posterior sample of $\Delta$ obtained using the four different kernels. The size of the mode near 423 days, which is of great scientific interest, differs substantially among the samplers.  In the second column of Fig.~\ref{tdc2}, we magnify this mode, superimposing a curve that represents the marginal posterior density of $\Delta$ based on twenty million posterior samples obtained with an oracle sampler\footnote{We use an MH within Gibbs sampler equipped with an independent Metropolis kernel \citep{tierney1994} that is invariant to $\pi_{11}$. The jumping rule for this kernel is {Uniform[400, 450]} with probability 0.1 and from   Uniform[1050, 1178.939] otherwise. We emphasize that this algorithm would not be feasible without prior knowledge of the size and location of the two posterior modes.} constructed with knowledge of both mode locations. The size and shape of the mode near 423 days obtained with RAM match the oracle sampler better than the other algorithms, which is an algorithmic confirmation of the reliability of RAM.
%, even if this secondary mode might not be  in this example. 

\section{Concluding remarks} \label{sec:conclusion}
We propose RAM both as an alternative to deal with multimodality, and as a newer strategy of forming acceptance probabilities. It can also be viewed as using negative temperature in annealing type of strategies, as Professor Art Owen recognized in his comments on an early version of our paper.  

More work is needed to extend RAM's applicability. In particular, we plan to compare the theoretical convergence rate of our algorithm to others, but this is difficult partially owing to  the intractable down-up jumping density, $q^{\textrm{DU}}$. Also, a better set of strategies for tuning RAM in various multimodal cases needs to be investigated. Different ways to encourage a down-up movement in density may exist, e.g., mixing anti-Langevin and Langevin algorithms or tempering with negative and positive temperature levels, both of which were suggested in a personal blog of Professor Christian P. Robert\footnote{https://xianblog.wordpress.com/2016/01/28/love-hate-metropolis-algorithm/}.  Another avenue for further improvement is to apply the ideas of the mode-jumping proposal and the delayed rejection method to RAM, e.g., allowing an asymmetric density function $q$ so that the downhill move encourages longer jumps than the uphill move does. Using this down-up idea to construct a global optimizer  is another possible  extension as the tempering idea is used for a statistical annealing. We invite interested readers to explore these possibilities.

\bigskip
\noindent{\Large\bf Supplementary materials}

\begin{description}
\item[Appendices:] Appendices A, B, and C as cited in the article (Appendices.pdf).

\item[R code and data:] All the R code and data used in this article (RAM.zip).
%\item[R codes:] .
\end{description}

\bigskip
\noindent{\Large\bf Acknowledgements}\\
This project was conducted under the auspices of the CHASC International Astrostatistics Center. CHASC is supported by the NSF grants DMS 1208791, 1209232,  1513484, 1513492, and  1513546. Xiao-Li Meng acknowledges partial financial support from the  NSF grants given to CHASC. Hyungsuk Tak acknowledges partial support from the NSF grant DMS 1127914 given to SAMSI.  David A. van Dyk acknowledges support from a Wolfson Research Merit Award (WM110023) provided by the British Royal Society and from a Marie-Curie Career Integration Grant (FP7-PEOPLE-2012-CIG-321865) provided by the European Commission.  The authors thank the associate editor and two referees for their insightful comments and suggestions that significantly improved the presentation. We also thank Christian~P.~Robert,  Pierre~E.~Jacob, Art~B.~Owen, and Natesh~S.~Pillai for very helpful discussions and Steven R. Finch for his proofreading. 

\bibliographystyle{apalike}

\bibliography{bibliography}

\newpage 

\begin{appendices}

\section{The average number of density evaluations in Section~3.1}
\cite{kou2006} implement the EE (equi-energy) sampler by running five parallel chains under five different temperature levels. The chain under the highest temperature adopts only MH transitions, and the other four chains use an EE jump with probability 0.1 and an MH transition otherwise at each iteration.  The EE sampler begins by running a chain under the highest temperature for 75,000 iterations; the first 25,000 are burn-in iterations and the next 50,000 iterations form an energy ring at the highest temperature. The first chain uses only MH transitions. After running the first chain for 75,000 iterations, the sampler initiates the next chain under the second highest temperature and runs it for 75,000 iterations; the first 25,000 are burn-in iterations and the next 50,000 iterations form an energy ring at the second highest temperature. From the second chain, the sampler adopts an EE jump with probability 0.1 and an MH transition otherwise at each iteration. This process continues until the EE sampler finishes running the fifth chain under the unit temperature for 75,000 iterations; the first 25,000 iterations are discarded. All chains keep running until the end of the fifth chain, which means that the first chain runs for $5\times$75,000 iterations in total and the second one runs for $4\times$75,000 iterations, etc.  Each EE jump needs to evaluate the target density twice  (with caching). Thus, the (expected) total number of the  density evaluations is $16\times$75,000 and that per iteration is 16.0.

Similarly, \cite{kou2006} implement  parallel tempering with five temperature levels and  propose four swaps with probability 0.1 at the end of each iteration. Five chains under five different temperature levels are run simultaneously for 75,000 iterations, using MH transitions. At the end of each iteration, four swaps occur with probability 0.1 and no swaps otherwise. Each swap requires two additional evaluations of the target (with caching). Thus, the (expected) total number of the target density evaluations is 435,000 and the average number of the  density evaluations per iteration is 5.8 (=435,000/75,000).

\section{Implementation details in Section 3.3}

Tempered transitions require several tuning parameters, e.g., the number of rungs of the temperature ladder and the temperature and jumping scale of each rung, and setting these parameters is  known to be challenging in practice \citep{behrens2012}. At each iteration, the tempered transitions ascend the temperature ladder to explore a flatter surface where the modes are melted down, and then descend the ladder, accepting the last candidate with a modified acceptance probability to maintain the stationary distribution (Neal, 1996). To sample $\pi_1$ in~(12) at iteration $i$, for example,  suppose ${\pi_{1j}(x_1)\propto  \{\pi_1(x_1\mid x_2^{(i-1)}, x_3^{(i-1)}, x_4^{(i-1)}, y, w)\}^{1/T_j}}$, where $T_j$ is the temperature at rung $j$ for $j=1, \ldots, J$. The target density is $\pi_{10}(x_1)$  and the ladder  has $J$ rungs with  {$T_0=1<T_1<\cdots<T_{J}$}.  Within each iteration~$i$, starting from $j=1$ to $J$, we generate $\hat{x}_{1j}$ from $\textrm{N}_2(\hat{x}_{1, j-1}, \Sigma_j)$, where $\hat{x}_{10}=x_1^{(i-1)}$, and accept it with probability $\min\{1, \pi_{1j}(\hat{x}_{1j})/\pi_{1j}(\hat{x}_{1, j-1})\}$ and  set $\hat{x}_{1j}=\hat{x}_{1, j-1}$ otherwise.  Once we reach $j=J$, we reverse the process from $j=J$ to $1$ and generate $\check{x}_{1, j-1}$ from  $\textrm{N}_2(\check{x}_{1j}, \Sigma_j)$ where $\check{x}_{1J}=\hat{x}_{1J}$, and accept it with probability $\min\{1, \pi_{1, j-1}(\check{x}_{1, j-1})/\pi_{1, j-1}(\check{x}_{1j})\}$ and set $\check{x}_{1, j-1}=\check{x}_{1j}$ otherwise until  we reach the bottom of the temperature ladder, collecting $\check{x}_{1, J-1}, \ldots, \check{x}_{10}$. After generating the last proposal $\check{x}_{10}$, we set $x_1^{(i)}=\check{x}_{10}$  with an MH acceptance probability of
\begin{equation}
\min\left\{1,~~ \frac{\pi_{11}(x_1^{(i-1)})}{\pi_{10}(x_1^{(i-1)})}\times\cdots\times\frac{\pi_{1J}(\hat{x}_{1, J-1})}{\pi_{1, J-1}(\hat{x}_{1, J-1})} \frac{\pi_{1, J-1}(\check{x}_{1, J-1})}{\pi_{1J}(\check{x}_{1, J-1})}\times\cdots\times\frac{\pi_{10}(\check{x}_{10})}{\pi_{11}(\check{x}_{10})}\right\},\nonumber
\end{equation}
and set $x_1^{(i)}=x_1^{(i-1)}$ otherwise.

To fit the simulated data, we set three rungs with temperature equal to $2^j$ for the $j$th rung. Because the longest observed distance between two sensors is about 0.9, we set the jumping covariance  matrix $\Sigma_j=(0.9\times 1.2^{j-1})^2\times I_2$ for each Metropolis update of the tempered transitions for the $j$th rung. For Metropolis and RAM, we set $\Sigma=1.08^2\times I_2$. This is the same as the jumping covariance matrix of tempered transitions at the middle rung, i.e., $\Sigma_2$. An initial value for each unknown location for each Markov chain is randomly selected from the unit square, $[0, 1]\times[0, 1]$.

\section{The average number of $\pi_{11}$ evaluations in Section~3.4}
Since the  kernels are used only to sample $\pi_{11}$ in \emph{Step}~1 of Table~5, the average number of  $\pi_{11}$ evaluations at each iteration ($\textrm{N}_{\pi_{11}}$) is not proportional to the entire CPU time needed for sampling the full posterior $\pi$ in (18).  For reference, $\textrm{N}^{\textrm{M}}_{\pi_{11}}=2$ (with either the Gaussian or mixture proposal), $\textrm{N}^{\textrm{RAM}}_{\pi_{11}}=8.76$ ($\textrm{N}_d=1$, $\textrm{N}_u=4.48$,  $\textrm{N}_z=1.28$), and $\textrm{N}^{\textrm{TT}}_{\pi_{11}}=11$. Each Metropolis transition evaluates $\pi_{11}$ twice, once for the current state and once for the proposal at each iteration. Caching the density value of the current state  does not help reduce $\textrm{N}_{\pi_{11}}$ because the density of the current state changes according to the updates of the other parameters in \emph{Steps} 2 and 3 of Table~5.

\end{appendices}

\end{document}